\documentclass[aps,prl,twocolumn,superscriptaddress]{revtex4-1}
\usepackage{graphicx}% Include figure files
\usepackage{dcolumn}% Align table columns on decimal point
\usepackage{bm}
\usepackage{amsfonts,amsmath}
\usepackage{xcolor}
\usepackage{float}

\usepackage[bookmarks=true,colorlinks,linkcolor=black]{hyperref}
\hypersetup{urlcolor=blue,citecolor=blue}                                        % This two orders change the color of links 
\usepackage{ulem}

\newcommand\PICorientpluscross{\setlength{\unitlength}{2mm}
\begin{picture}(2.9,2)(-1.4,-0.6)
\put(-1,-1){\vector(1,1){2.3}}%
\put(-1,1){\vector(-1,1){0.3}}%
\qbezier(-0.3,0.3)(-0.8,0.8)(-1.0,1.0)
\qbezier(0.3,-0.3)(0.8,-0.8)(1.0,-1.0)
\end{picture}}

\newcommand\PICorientminuscross{\setlength{\unitlength}{2mm}
\begin{picture}(2.9,2)(-1.4,-0.6)
\put(1,-1){\vector(-1,1){2.3}}%
\put(1,1){\vector(1,1){0.3}}%
\qbezier(0.3,0.3)(0.8,0.8)(1.0,1.0)
\qbezier(-0.3,-0.3)(-0.8,-0.8)(-1.0,-1.0)
\end{picture}}

\begin{document}

%\citet

%\preprint{}

\title{Knot theory for two-band model of two-dimensional square lattice with high topological numbers}


\author{Xin LIU}
%\thanks{}
\affiliation{Beijing-Dublin
International College, Beijing University of Technology, Beijing 100124,
P.R. China}
\affiliation{Institute of Theoretical Physics, Beijing University of
Technology, Beijing 100124, P.R. China}
\author{Zhiwen CHANG}
\affiliation{Institute of Theoretical Physics, Beijing University of
Technology, Beijing 100124, P.R. China}
\author{Weichang HAO}
\email{whao@buaa.edu.cn}
%\thanks{whao@buaa.edu.cn}
\affiliation{School of Physics, Beihang University, Beijing 100191, P.R. China}


\date{\today}

\begin{abstract}
A knot theory for two-dimensional square lattice is proposed, which sheds light on design of new two-dimensional material with high topological numbers. We consider a two-band model, focusing on the Hall conductance $\sigma_{xy} = \frac{e^2}{h}P$, where $P$ is a topological number, the so-called Pontrjagin index. By re-interpreting the periodic momentum components $k_x$ and $k_y$ as the string parameters of two entangled knots, we discover that $P$ becomes the Gauss linking number between the knots. This leads to a successful re-derivation of the typical $P$-evaluations in literature:  $P = 0, \pm 1$. Furthermore, with the aid of this explicit knot theoretical picture we modify the two-band model to achieve higher topological numbers, $P = 0, \pm 1, \pm 2$.
\end{abstract}

%\keywords{}

\maketitle


Topological insulators (TIs) have experienced a rapid development ever since the experimental observation of the quantum spin Hall effect in HgTe quantum wells \cite{10.1126/science.1133734,10.1126/science.1148047} and the quantum anomalous Hall effect (QAHE) in chromium-doped (Bi,Sb)$_{2}$Te$_3$ \cite{10.1126/science.1234414}. A TI has gapless edge states which arise from the band structure and are characterized by a quantized topological number; topology-protected edge states are insensitive to disorder due to absence of backscattering states. Electron-electron interactions do not modify the edge states \cite{10.1103/RevModPhys.82.3045}, so TIs are predicted to have properties useful for designing spintronics devices and quantum computers \cite{10.1103/RevModPhys.82.3045,10.1103/RevModPhys.83.1057,10.1103/PhysRevLett.100.156404}.

The tight-binding model for an $N$-band insulator is given by
$
H=\sum_{\mathbf{k};\alpha,\beta}\psi^{\alpha\ast}_{\mathbf{k}}h_{\mathbf{k}}^{\alpha\beta}\psi_{\mathbf{k}}^{\beta}
% \label{eq:201908065}
$,
where $\mathbf{k}=\left(k_x , k_y \right)$ is the two-dimensional momentum (for convenience the subscript $\mathbf{k}$ being ignored below). $\alpha, \beta=1, 2, \cdots, N$ are the band indices; if specially $N=2$, it is a $2$-band structure, the minimal model to have nontrivial topology. The first $2$-band example with nonzero topological number was proposed by Haldane in the honeycomb lattice \cite{10.1103/PhysRevLett.61.2015} which realizes QAHE as a magnetic topological insulator with breaking time-reversal symmetry. Recently the $3$-band model in a Kagome or Lieb lattice attracted much interest \cite{10.1103/PhysRevLett.106.236802,10.1103/PhysRevB.82.085310}. In this paper we focus on the $2$-band model in a square lattice; the method developed could expectedly be generalized to a $3$-band study.

%%%%%%%%%%%%%%%%%%%%%%%%%%%%%%%%%%%%%%%%%%%%%%%%%
%%%%%%%%%%%%%%%%%%%%%%%%%%%%%%%%%%%%%%%%%%%%%%%%%
%%%%%%%%%%%%%%%%%%%%%%%%%%%%%%%%%%%%%%%%%%%%%%%%%

\textit{Pontrjagin topological index.} Introducing a Weyl spinor
$\Psi=\left(\begin{array}{cc}
 \psi^{1} &  \psi^{2} \end{array} \right)^T$,
the Hamiltonian above becomes
$
H=\Psi^{\dagger}h\Psi   %\label{eq:201908091}
$.
Expanding $h$ onto the Clifford algebraic basis $\left(I, \boldsymbol{\sigma}\right)$, with $\sigma^{a}$, $a=x,y,z$ the Pauli matrices, we have
$
h= \epsilon I+\sum_{a=x,y,z}h_{a}\sigma^{a}%, \label{eq:201908093}
$. The $3$-vector $\boldsymbol{h} = \left(h_x,h_y,h_z \right)$ induces a unit vector, $\hat{\boldsymbol{h}} = \frac{\boldsymbol{h}}{\left| \boldsymbol{h} \right|} $, which forms a unit $2$-sphere $S^2$ in the $SU(2)$ group space. Due to the periodicity $k_{x}, k_{y}\in \left[0,2\pi \right)$, $\hat{\boldsymbol{h}}$ defines a topological map:
$
\hat{\boldsymbol{h}} (\mathbf{k}): S^{1}\times S^{1} \rightarrow S^2 
%\label{topomapS1xS1toS2}
$, where the first Brillouin zone $S^{1}\times S^{1}$ is treated as a two-dimensional surface of torus $T^2$, with the ordered pair $\left(k_x , k_y \right) \in S^{1}\times S^{1}$ represented by a point on the torus.

Physical properties of QAHE in this system \cite{10.1103/PhysRevB.78.195424,10.1103/PhysRevB.74.085308} arise from the topological degree of this $\hat{\boldsymbol{h}}$-map, which is an integer-valued topological invariant known as the Pontrjagin index \cite{10.1017/CBO9781139015509},
\begin{equation}
P = \frac{1}{4 \pi}\int_{S^{1}\times S^{1}} \hat{\boldsymbol{h}}\cdot \left(\partial_{k_{x}}\hat{\boldsymbol{h}} \times \partial_{k_{y}}\hat{\boldsymbol{h}}\right) dk_x \wedge dk_y. \label{eq:2019080920}
\end{equation}
$P$ leads to the quantized Hall conductance, $\sigma_{xy} = \frac{e^2}{h}P$. The integrand of eq.\eqref{eq:2019080920} carries the geometric meaning of a solid angle. In the $SU(2)$ group space, when performed on the topologically non-trivial sphere $S^2$, the integral $P$ describes the total topological charge of monopoles (three-dimensional point defects) occurring at $\boldsymbol{h} = \boldsymbol{0}$; when performed on the north and south hemispheres separately, $P$ describes the topological charges of merons (two-dimensional point defects) occurring at $h_x = h_y =0$ and $h_z \neq 0$, because a hemisphere is diffeomorphic to a topologically-trivial Euclidean two-dimensional disk. For the latter meron case, $P$ is identical to a first-Chern number, since the integrand of eq.\eqref{eq:2019080920} can be turned into a $U(1)$ gauge field tensor serving as a first-Chern class: 
$\hat{\boldsymbol{h}}\cdot \left(\partial_{\mu}\hat{\boldsymbol{h}} \times \partial_{\nu}\hat{\boldsymbol{h}}\right)
= \partial_{\mu} W_{\nu} -  \partial_{\nu} W_{\mu}
%\label{Wu-YangPotential-1}
$. Here $W_{\mu},~\mu=k_x,k_y$, is the so-called Wu-Yang potential \cite{10.1103/PhysRevD.12.3845}  given by $
W_{\mu} = \hat{\boldsymbol{e}} \cdot \partial_{\mu} \hat{\boldsymbol{f}}
$, with $\hat{\boldsymbol{e}}$ and $\hat{\boldsymbol{f}}$ being two perpendicular unit vectors on the $\hat{\boldsymbol{h}}$-formed $S^2$ (i.e., $\left\{ \hat{\boldsymbol{h}}, \hat{\boldsymbol{e}} , \hat{\boldsymbol{f}}\right\}$ forms an orthonormal frame). The locations of the defects give the corresponding Dirac points where edge states take place.

In this paper we focus on the topological number $P$.
The usual way to obtain it is to perform direct computation of the integral \eqref{eq:2019080920}, as long as a concrete two-band model, like eq.\eqref{eq:201908011} below, is provided \cite{10.1103/PhysRevB.78.195424}.
The disadvantage of this method is the absence of an explicit geometric illustration of the expression \eqref{eq:2019080920}. It hinders a direct read-out of the $P$ value, no mentioning further manipulation and design of new material with higher complexity in topology. We need an alternative picture beyond the point representation to present a clear understanding of the problem. For this purpose let us introduce a new vector representation for $\left(k_x , k_y \right)$ and the induced Gauss mapping of knot theory.

%%%%%%%%%%%%%%%%%%%%%%%%%%%%%%%%%%%%%%%%%%%%%%%
%%%%%%%%%%%%%%%%%%%%%%%%%%%%%%%%%%%%%%%%%%%%%%%
%%%%%%%%%%%%%%%%%%%%%%%%%%%%%%%%%%%%%%%%%%%%%%%

\begin{figure}[H]
\centering \includegraphics[width=0.43\textwidth]{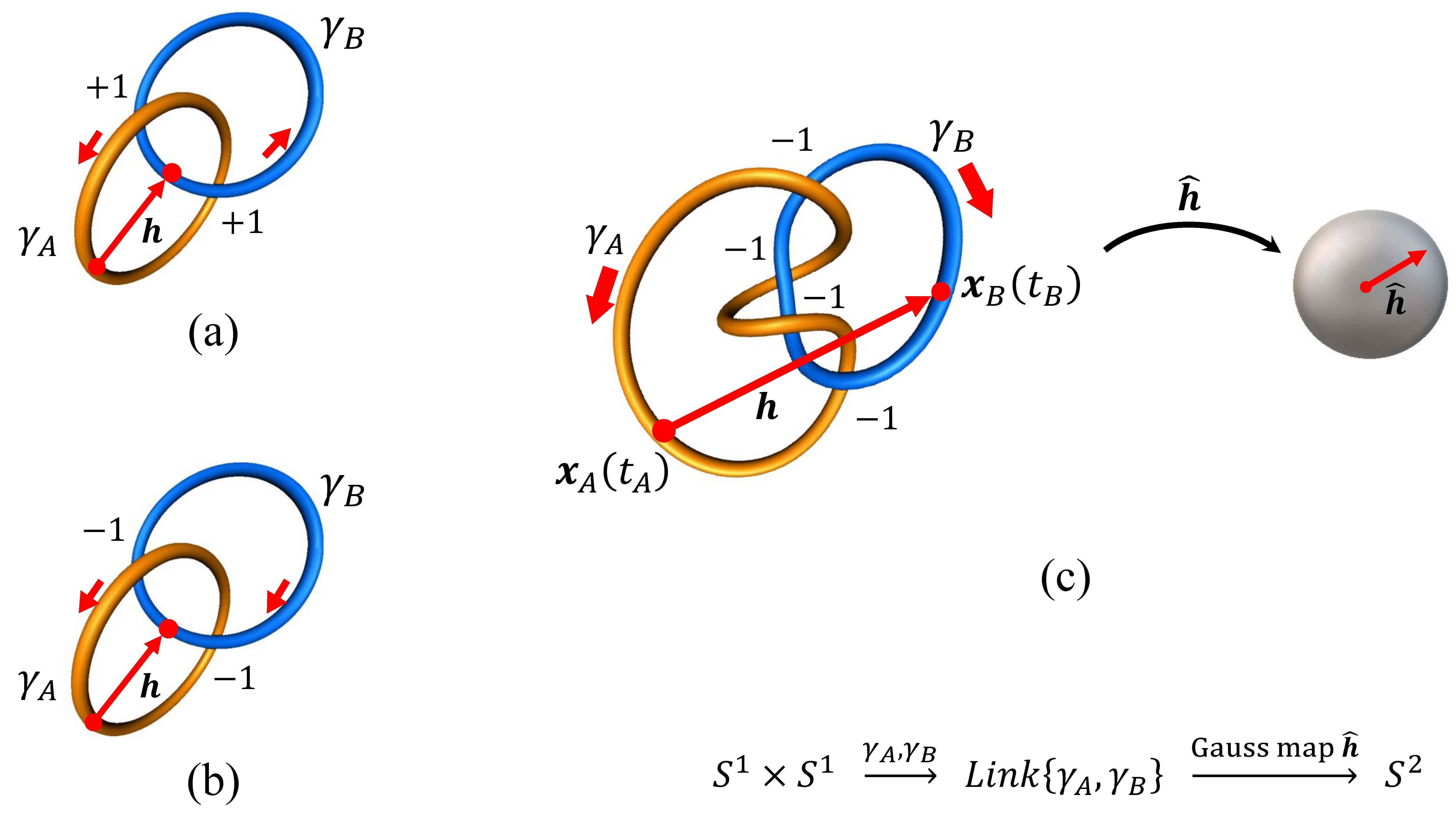}
\caption{Three links with dual knot components, $\left\{ \gamma_A,\gamma_B \right\}$: (a) linking number $+1$; (b) linking number $-1$; (c) linking number $-2$. The $\boldsymbol{x}_A$ and $\boldsymbol{x}_B$ are two points picked from $\gamma_A$ and $\gamma_B$, respectively. The unit vector $\hat{\boldsymbol{h}} = \frac{\boldsymbol{h}}{\left| {\boldsymbol{h}} \right|}$, defined from $\boldsymbol{h}  = \boldsymbol{x}_B  - \boldsymbol{x}_A $, gives a Gauss map.
\label{Fig1-LinkNum-S2}}
\end{figure}

\textit{Vector representation for $\left( k_x , k_y  \right)$, Gauss mapping and linking number.} Consider a link $\left\{ \gamma_A,\gamma_B \right\}$ in Figure \ref{Fig1-LinkNum-S2}(a), (b) or (c), where $\gamma_A$ and $\gamma_B$ are two knots. 

\noindent Let $\boldsymbol{x}_A=\boldsymbol{x}_A \left( t_A \right)$ and $\boldsymbol{x}_B=\boldsymbol{x}_B \left( t_B \right)$ be two arbitrary points picked from $\gamma_A$ and $\gamma_B$, respectively, with $t_A$ and $t_B$ being two periodic string parameters, $t_A, t_B \in \left[0,2\pi\right)$, i.e., $\left\{ t_A,t_B \right\} \in S^1 \times S^1$. Introducing a vector $\boldsymbol{h}  = \boldsymbol{x}_B - \boldsymbol{x}_A$, one can define a unit vector $\hat{\boldsymbol{h}} = \frac{\boldsymbol{h}}{\left| {\boldsymbol{h}} \right|}$ which gives a Gauss map $\hat{\boldsymbol{h}}:\left\{ \gamma_A,\gamma_B \right\} \rightarrow S^2$. We have the mapping:
$S^{1}\times S^{1} \overset{\gamma_A, \gamma_B}{\longrightarrow}  \left\{\gamma_A, \gamma_B\right\} 
\overset{\hat{\boldsymbol{h}}}{\longrightarrow}  S^2
$. That means, when $t_A$ and $t_B$ run out the two $S^1$'s once, $\boldsymbol{x}_A$ and $\boldsymbol{x}_B$ run out $\gamma_A$ and $\gamma_B$ once, respectively, such that $\hat{\boldsymbol{h}}$ covers the unit sphere $S^2$ for $\mathrm{Deg} \left( \hat{\boldsymbol{h}} \right)$ times. Here the Gauss mapping degree is defined as \cite{10.1098/rspl.1830.0095,10.1142/S0218216511009261}
\begin{eqnarray}
&& \mathrm{Deg} \left( \hat{\boldsymbol{h}} \right)
= \hat{\boldsymbol{h}}^{*}\left( \frac{1}{4 \pi}\int_{S^2} \hat{\boldsymbol{h}} \cdot d\hat{\boldsymbol{h}} \times d\hat{\boldsymbol{h}} \right) \notag \\
&=& \frac{1}{4 \pi}\int_{S^{1}\times S^{1}} \hat{\boldsymbol{h}}\cdot \left(\partial_{t_{A}}\hat{\boldsymbol{h}} \times \partial_{t_{B}}\hat{\boldsymbol{h}}\right)
dt_A \wedge dt_B ,
\label{GaussMapDegLk}
\end{eqnarray}
where $ \hat{\boldsymbol{h}}^{*}\left( \bullet \right)$ represents a pull-back of the $ \hat{\boldsymbol{h}}$-map.
Eq.\eqref{GaussMapDegLk} is recognized to be the same as the Pontrjagin index \eqref{eq:2019080920} up to a difference in the parameters $\left(t_A , t_B \right)$ and $\left( k_x , k_y \right)$.

In knot theory it is known $\mathrm{Deg} \left( \hat{\boldsymbol{h}} \right)$ is equal to the Gauss mutual linking number between $\gamma_A$ and $\gamma_B$ \cite{10.1098/rspl.1830.0095,10.1142/S0218216511009261}:
\begin{eqnarray}
&& Lk \left( \gamma_A,\gamma_B \right) = \oint_{\gamma_A}\oint_{\gamma_B} \frac{\left( \boldsymbol{x}_B - \boldsymbol{x}_A \right)
\cdot \left( d\boldsymbol{x}_A \times d\boldsymbol{x}_B \right) }{\| \boldsymbol{x}_B - \boldsymbol{x}_A \|^3} \notag \\
&=& \int_{0}^{2\pi}\int_{0}^{2\pi} \frac{\left( \boldsymbol{x}_B  - \boldsymbol{x}_A  \right)
\cdot \left( \dot{\boldsymbol{x}}_A \times \dot{\boldsymbol{x}}_B \right) }{\| \boldsymbol{x}_B - \boldsymbol{x}_A \|^3}
dt_A \wedge dt_B .
\label{eq:201908017}
\end{eqnarray}
In practice the linking number $Lk \left( \gamma_A,\gamma_B \right)$ can be computed using a much easier algebraic method instead of the complicated integrals \eqref{eq:201908017} and \eqref{GaussMapDegLk}:
\begin{equation}
Lk \left( \gamma_A,\gamma_B \right) = \frac12 \sum\limits_{r=1}^{n} \epsilon_r, \label{LinkAlgeCount}
\end{equation}
where $n$ denotes the total number of mutual crossing sites between $\gamma_A$ and $\gamma_B$, with $\epsilon_r$ the algebraic degree of the $r$th site. Here the degree $\epsilon$ is defined as: $\epsilon = +1$ for \PICorientpluscross ; $\epsilon = -1 $ for \PICorientminuscross . Typical examples are shown in Figure \ref{Fig1-LinkNum-S2}. Thus, when facing a link, one can simply count the algebraic degree of every single mutual crossing site, and then sum all them up to obtain the total linking number.

Now we are at the stage to substitute $k_x$ and $k_y$ into $t_A$ and $t_B$, respectively. An important fact is: the condition of doing this substitution is that the expression of $\boldsymbol{h}$ in a given model can be clearly separated up into a pure $k_x$ part and a pure $k_y$ part:
  \begin{equation}
\boldsymbol{h} \left( k_x , k_y\right) = \boldsymbol{x}_B \left( k_y \right) - \boldsymbol{x}_A \left( k_x \right).
\label{hkxky-Def}
\end{equation}
If this condition is satisfied, the Pontrjagin index $P$ achieves a Gauss linking number realization:
\begin{equation}
P = \mathrm{Deg} \left(\hat{\boldsymbol{h}}\right)
= Lk \left[ \gamma_A \left( k_x \right),\gamma_B \left( k_y \right) \right] = \frac12 \sum\limits_{r=1}^{n} \epsilon_r.
\end{equation}
The above acts as a knot theoretical method, with $\boldsymbol{h}$ being a vector representation for the ordered pair $\left( k_x , k_y  \right)$.

%%%%%%%%%%%%%%%%%%%%%%%%%%%%%%%%%%%%%%%%%%%%%%%%%
%%%%%%%%%%%%%%%%%%%%%%%%%%%%%%%%%%%%%%%%%%%%%%%%%
%%%%%%%%%%%%%%%%%%%%%%%%%%%%%%%%%%%%%%%%%%%%%%%%%

\textit{Example to test the method.} Let us check a typical two-band model in literature \cite{10.1103/PhysRevB.78.195424}. Consider a square lattice generated by perpendicular primary vectors, $\boldsymbol{a}_{1}=(1,0)l$ and $\boldsymbol{a}_{2}=(0,1)l$, with $l=1$ the lattice constant; here only the nearest neighboring (NN) interactions are involved. The Hamiltonian $\boldsymbol{h} = \left( h_x, h_y, h_z \right)$ is given by
\begin{equation}
h_{x}=\sin k_{x}, ~~
h_{y}=\sin k_{y},~~
h_{z}=m+\cos k_{x}+\cos k_{y}, \label{eq:201908011}
\end{equation}
where $m \in \mathbb{R}$ is an on-site energy to open up an energy gap. The Pontrjagin index $P$ in this case takes various values due to varying $m$ \cite{10.1103/PhysRevB.78.195424}:
\begin{equation}
P =
   \left\{
       \begin{array}{lcl}
                     0, &  \text{when\quad}  & m>2 \hspace{2mm} \textrm{or}\hspace{2mm} m<-2; \\
                      -1, &  \text{when\quad}  & 0<m<2; \\
                                         +1,  & \text{when\quad} & -2<m<0 , \\
                                         \text{indeterminate}, & \text{when\quad}  & m=0,\pm 2 .
         \end{array}
  \right.  \label{eq:201908012}
\end{equation}
The Dirac points are $\left(k_x , k_y \right) = \left(0 , 0 \right)$, $\left(0 , \pi \right)$, $\left(\pi , 0 \right)$ and $\left(\pi , \pi \right)$. At those points, if $m=-2,0,0,2$, respectively, one has $h_x = h_y = h_z = 0$, and the monopole defects occur; otherwise, if $m$ does not take these values there, one has  $h_x = h_y = 0$ and $h_z \neq 0$, and the merons occur.

\begin{figure}[H]
\centering \includegraphics[width=0.43\textwidth]{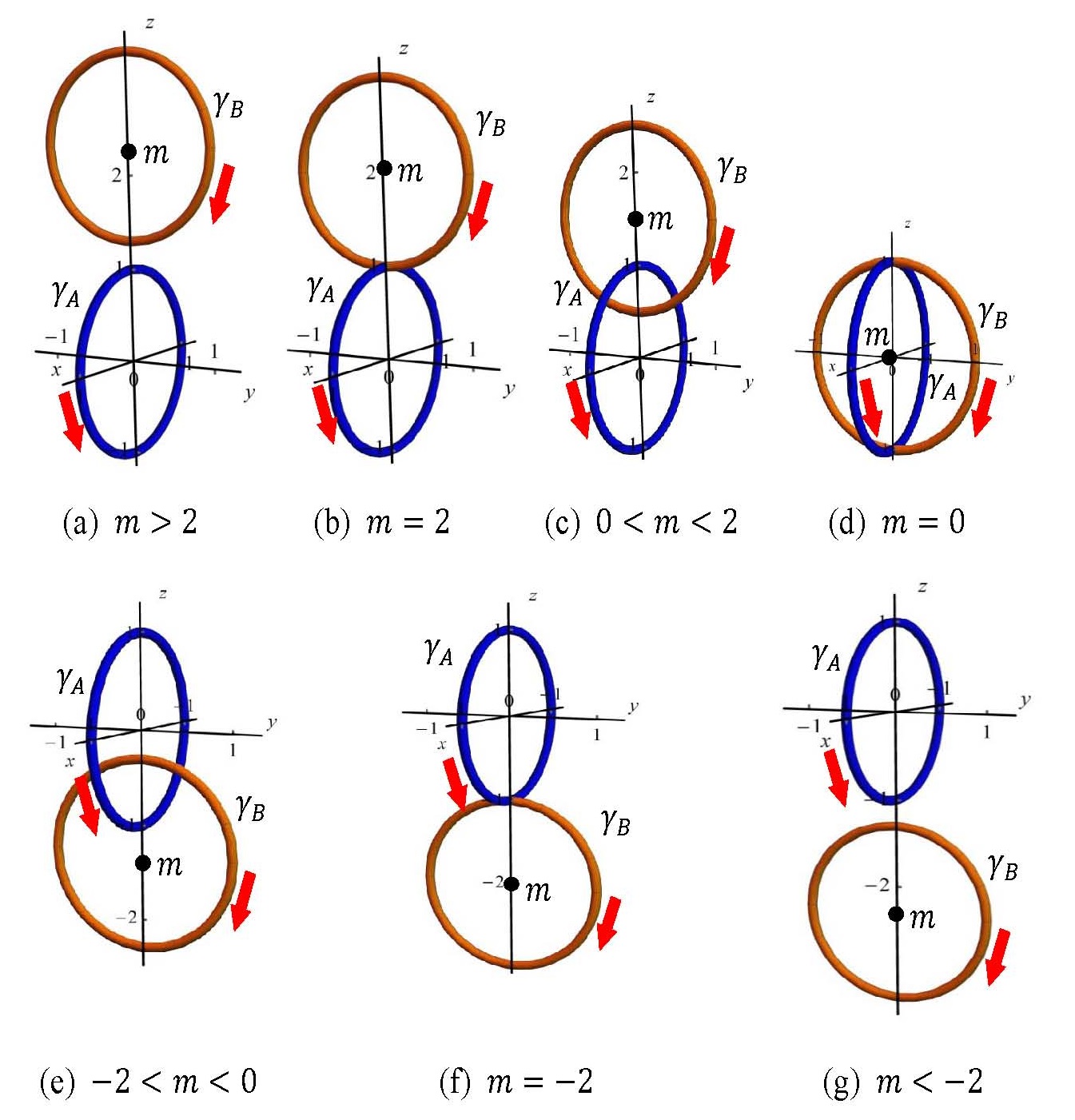}
\caption{$\gamma_A$ is a unit circle in the $xz$-plane, centered at $(0,0,0)$; $\gamma_B$ is a unit circle in the $yz$-plane, centered at $(0,0,m)$. Cases (a)--(g) show the different relevant positions of $\gamma_{A}$ and $\gamma_{B} $, corresponding to various linkage situations: (a) and (g), $\gamma_{A}$ and $\gamma_{B}$ are disjoint, hence $Lk \left(\gamma_{A},\gamma_{B} \right)=0$; (b), (d) and (f), $\gamma_{A}$ and $\gamma_{B}$ contact, hence $Lk \left(\gamma_{A},\gamma_{B} \right)$ is indeterminate; (c), $Lk \left(\gamma_{A},\gamma_{B} \right)=-1$; (e), $Lk \left(\gamma_{A},\gamma_{B} \right)=+1$.
\label{Fig3-Cases-a_to_g}}
\end{figure}

Now let us rewrite $\left(h_x,h_y,h_z \right)$ separately into a pure $k_x$ part and a pure $k_y$ part as per eq.\eqref{hkxky-Def} :
\begin{equation}
\gamma_{A}:
         \left\{
               \begin{array}{l}
                      x_{A} \left( k_x\right)=-\sin k_{x},  \\
                      y_{A} \left( k_x\right)=0,   \\
                      z_{A} \left( k_x\right)=-\cos k_{x};
              \end{array}
       \right.
~~
\gamma_{B}:
         \left\{
               \begin{array}{l}
                      x_{B}  \left( k_y\right)=0,  \\
                      y_{B}  \left( k_y\right)= \sin k_{y},   \\
                      z_{B}  \left( k_y\right)= \cos k_{y} + m.
              \end{array}
       \right.
\label{eq:201907285}
\end{equation}
Obviously $h_x = x_B - x_A$, $h_y = y_B - y_A$ and $h_z = z_B - z_A$, while $\left(x_A,y_A,z_A\right)$ and $\left(x_B,y_B,z_B\right)$ form two unit circles $\gamma_A$ and $\gamma_B$ in the $xz$- and $yz$-planes, respectively, as shown in Figure \ref{Fig3-Cases-a_to_g}(a)--(g).

\vspace*{10mm}

\noindent When $k_x$ and $k_y$ increase from $0$ to $2\pi$, $\gamma_A$ and $\gamma_B$ obtain their respective orientations. The varying $m$ leads to different relevant positions of $\gamma_{A}$ and $\gamma_{B}$, and therefore various linkage situations:
\begin{itemize}

%\itemsep=1pt
%\parsep=1pt
%\parskip=1pt

\item When $m>2$ or $m<-2$, $\gamma_{A}$ and $\gamma_{B}$ are apart from each other, hence the linking number $Lk \left(\gamma_{A},\gamma_{B} \right)=0$, corresponding to Figure \ref{Fig3-Cases-a_to_g}(a) and (g).

\item When $m=0,\pm 2$, $\gamma_{A}$ and $\gamma_{B}$ contact, hence the linking number is indeterminate, corresponding to Figure \ref{Fig3-Cases-a_to_g}(b), (d) and (f).

\item When $0<m<2$, $\gamma_{A}$ and $\gamma_{B}$ has linking number $Lk \left(\gamma_{A},\gamma_{B} \right)=-1$, corresponding to Figure \ref{Fig3-Cases-a_to_g}(c). This case is similar as Figure \ref{Fig1-LinkNum-S2}(b).

\item When $-2<m<0$, $\gamma_{A}$ and $\gamma_{B}$ has linking number $Lk \left(\gamma_{A},\gamma_{B} \right)=+1$, corresponding to Figure \ref{Fig3-Cases-a_to_g}(e). This case is similar as Figure \ref{Fig1-LinkNum-S2}(a).

\end{itemize}
These cases and Figure \ref{Fig3-Cases-a_to_g}(a)--(g) precisely reproduce the different evaluations of the Pontrjagin index $P$ in eq.\eqref{eq:201908012}.

%%%%%%%%%%%%%%%%%%%%%%%%%%%%%%%%%%%%%%%%%%%
%%%%%%%%%%%%%%% New section %%%%%%%%%%%%%%%%%%%%%
%%%%%%%%%%%%%%%%%%%%%%%%%%%%%%%%%%%%%%%%%%%
%\newpage

\textit{Modified two-band model to realize higher topological numbers.}
Next we propose a modified two-band model to achieve higher Pontrjagin indices, i.e., $P=Lk \left( \gamma_A , \gamma_B\right) = 0,\pm 1, \pm 2$. The high topological numbers \cite{10.1103/PhysRevLett.111.136801} might contribute to effectively reduce contact-resistance and significantly improve performance of interconnect devices, within dissipationless conduction of edge channels in a quantum Hall insulator. In the simulation of Zhang {\it et al.} \cite{10.1103/PhysRevLett.111.136801}, high topological number plateaus are expectedly achievable from increasing the magnetic doping concentration in Cr-doped Bi$_2$(Se,Te)$_3$. Unfortunately, this proposal faces a great challenge in material growth.  In contrast, 
our strategy is to place emphasis on intrinsic lattice symmetry to obtain
high topological numbers.

Our proposal is to generalize eq.\eqref{eq:201907285} and Figure \ref{Fig3-Cases-a_to_g} to Figure \ref{Fig3-mod2band} and eqs.\eqref{Linknum2-gammaA} and \eqref{Linknum2-gammaB} below. In this modified model an important requirement is to set up the domains of the momentum components to be $k_x \in  \left[0,2\pi \right)$ and $k_y \in  \left[0,4\pi \right)$. The augmented domain $\left[0,4\pi \right)$ will be treated as two separate branches $\left[0,2\pi \right)$ and $\left[2\pi , 4\pi \right)$. See below.

\begin{figure}[H]
 \centering
   \includegraphics[width=0.12\textwidth]{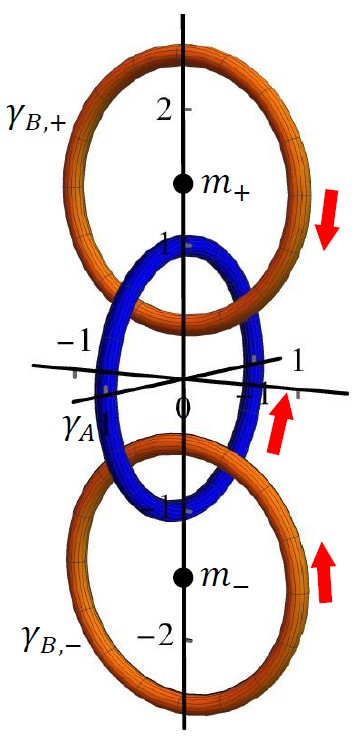}
\caption{$\gamma_A$ is a unit circle in the $xz$-plane, centered at $(0,0,0)$. $\gamma_B$ is composed of two branches, i.e., two unit circles in the $yz$-plane: $\gamma_{B,+}$, centered at $(0,0,m_+)$, with $k_y \in \left[0,2\pi \right)$;  $\gamma_{B,-}$, centered at $(0,0,m_-)$, with $k_y \in \left[2\pi , 4\pi \right)$. The orientations of the circles are as shown. When $m_+$ and $m_-$ take varying values, the linking number $Lk \left(\gamma_{A},\gamma_{B} \right)$ might have various values $0, \pm 1, \pm 2$. The parameters taken here are: $m_+=1.3$ and $m_-=-1.3$, which yield $Lk \left(\gamma_{A},\gamma_{B} \right)=-2$.
\label{Fig3-mod2band}}
\end{figure}

The respective expressions for $\gamma_A$ and $\gamma_B$ read
\begin{itemize}

\item $\gamma_A$: The same as in eq.\eqref{eq:201907285}, with $k_x \in \left[ 0,2\pi \right)$.
\begin{equation}
         \gamma_A: \left\{
               \begin{array}{l}
                      x_{A} \left( k_x\right)=-\sin k_{x},  \\
                      y_{A} \left( k_x\right)=0,   \\
                      z_{A} \left( k_x\right)=-\cos k_{x},
              \end{array}
       \right.
\label{Linknum2-gammaA}
\end{equation}

\item $\gamma_B$: The period of $k_{y}$ is augmented to $\left[0,4\pi \right)$.
    \begin{equation}
    \begin{array}{lll}
\gamma_{B,+}:
 \left\{
      \begin{array}{l}
                 x_{B,+}  \left( k_y\right) = 0,\\
                 y_{B,+}  \left( k_y\right) = \sin k_{y},\\
                 z_{B,+}  \left( k_y\right) = \cos k_{y} + m_{+},
       \end{array}
 \right. & k_y \in \left[ 0, 2\pi \right);
\\ ~ \\
\gamma_{B,-}:
 \left\{
          \begin{array}{l}
                  x_{B,-} \left( k_y\right)  = 0,\\
                  y_{B,-} \left( k_y\right)  =  \sin\left(-k_{y}\right),\\
                  z_{B,-} \left( k_y\right)  =  \cos\left(-k_{y}\right) + m_{-},
          \end{array}
  \right. & k_y \in \left[ 2\pi, 4\pi \right).
 \label{Linknum2-gammaB}
\end{array}
\end{equation}
\end{itemize}
The ring $\gamma_{B,+}$ is centered at $\left(0,0,m_{+} \right)$, with clockwise rotation; $\gamma_{B,-}$ centered at $\left(0,0,m_{-} \right)$, with anticlockwise rotation. 

This modified model is able to produce higher topological Pontrjagin index $P = 0, \pm 1, \pm 2$. For instance, if specially letting the two rings $\gamma_{B,+}$ and $\gamma_{B,-}$ contact at one single point, denoted as $\left(0,0, m_0 \right)$, with $m_0$ an introduced parameter, $\gamma_{B}$ turns to form a figure-$8$ shape. And then the linkage $Lk \left(\gamma_{A},\gamma_{B} \right)$ is similar as in Figure \ref{Fig1-LinkNum-S2}(c). The figure-$8$ shape is realized by setting a constraint $\left\{
\begin{array}{l}
m_+ = m_0 +1; \\
m_- = m_0 -1, \\
\end{array}
\right.$ or 
$\left\{
\begin{array}{l}
m_- = m_0 +1; \\
m_+ = m_0 -1. \\
\end{array}
\right.$ In detail,
\begin{itemize}

\item for $\left\{
\begin{array}{l}
m_+ = m_0 +1; \\
m_- = m_0 -1, \\
\end{array}
\right.$ which means an upper $\gamma_{B,+}$ and a lower $\gamma_{B,-}$, we have
\begin{equation}
P =
\left\{
\begin{array}{lcl}
0, &  \text{when\quad}  & m_0>3; \\
 +1, &  \text{when\quad}  & 1<m_0<3; \\
-2, &  \text{when\quad}  & -1<m_0<1; \\
+1,  & \text{when\quad} & -3<m_0<-1; \\
0, &  \text{when\quad}  & m_0<-3; \\
\text{indeterminate}, & \text{when\quad}  & m_0=\pm 1, \pm 3 .
         \end{array}
  \right.  \label{eq:newmodelP-1}
\end{equation}

\item for $\left\{
\begin{array}{l}
m_- = m_0 +1; \\
m_+ = m_0 -1, \\
\end{array}
\right.$ which means an upper $\gamma_{B,-}$ and a lower $\gamma_{B,+}$, we have
\begin{equation}
P =
\left\{
\begin{array}{lcl}
0, &  \text{when\quad}  & m_0>3; \\
 -1, &  \text{when\quad}  & 1<m_0<3; \\
+2, &  \text{when\quad}  & -1<m_0<1; \\
-1,  & \text{when\quad} & -3<m_0<-1; \\
0, &  \text{when\quad}  & m_0<-3; \\
\text{indeterminate}, & \text{when\quad}  & m_0=\pm 1, \pm 3 .
         \end{array}
  \right.  \label{eq:newmodelP-2}
\end{equation}

\end{itemize} 

In this modified model, the first Brillouin zone is $\left(k_x , k_y \right) \in \left[0,2\pi \right) \times \left[0,4\pi \right)$. To illustrate this zone let us consider the $\frac{\mathbf{k}}{2}$-space. Regarding $\frac{k_x}{2}$ as the polar angle $\theta$ with period $\pi$, and $\frac{k_y}{2}$ as the azimuthal angle $\phi$ with period $2\pi$, we see the $\frac{\mathbf{k}}{2}$-space is a $2$-sphere $S^2$ actually. Hence the map $\frac{\mathbf{k}}{2} \rightarrow \hat{\boldsymbol{h}}\left(\frac{\mathbf{k}}{2} \right)$ becomes an $S^2 \rightarrow S^2$ map, in contrast with the original map $T^2 \rightarrow S^2$.

To realize the modified model in physics, we propose the following Hamiltonian:
in the momentum space,
\begin{eqnarray}
h &=&\sin k_{x}\sigma _{x}+\cos k_{x}\sigma _{z}  \nonumber \\
&& + H_{k_{y},\text{I}} \left\{ t_{+}\left[ \sin
k_{y}\sigma _{y}+\cos k_{y}\sigma _{z}\right] +m_{+}\sigma _{z}\right\}
\label{mod2band-m} \\
&&+H_{k_{y},\text{II}}
\left\{ t_{-}\left[ \sin \left( -k_{y}\right) \sigma _{y}+\cos \left(
-k_{y}\right) \sigma _{z}\right] +m_{-}\sigma _{z}\right\} , \nonumber
\end{eqnarray}%
where $t_+$ and $t_-$ are two adjustment parameters. Usually we take $t_+ = t_- =1$ as in eq.\eqref{Linknum2-gammaB}.
$H_{k_{y},\text{I}}  $ and $H_{k_{y},\text{II}}  $\ are two Heaviside-like stepwise
functions: when $k_{y}\in %
\left[ 0,2\pi \right) $, $H_{k_{y},\text{I}}  =1$ and $H_{k_{y},\text{II}} =0$; when $k_{y}\in \left[ 2\pi ,4\pi \right) $, $H_{k_{y},\text{I}}  =0$ and $H_{k_{y},\text{II}} =1$. In the real space,
\begin{eqnarray}
H &=&\sum_{n}t_0\left[ c_{n}^{\dagger }\frac{\sigma _{x}}{2i}c_{n+\boldsymbol{a}%
_{1}}+c_{n}^{\dagger }\frac{\sigma _{z}}{2}c_{n+\boldsymbol{a}_{1}}+h.c.%
\right]   \nonumber \\
&& + t_0\left[ c_{n-\boldsymbol{a}_{2}}^{\dagger }\frac{\sigma _{x}}{2i}c_{n-\boldsymbol{a}_{2}+\boldsymbol{a}%
_{1}}+c_{n-\boldsymbol{a}_{2}}^{\dagger }\frac{\sigma _{z}}{2}c_{n-\boldsymbol{a}_{2}+\boldsymbol{a}_{1}}+h.c.%
\right]   \nonumber \\
&&+t_{+}\left[ c_{n}^{\dagger }\frac{\sigma _{y}}{2i}c_{n+\boldsymbol{a}%
_{2}}+c_{n}^{\dagger }\frac{\sigma _{z}}{2}c_{n+\boldsymbol{a}_{2}}+h.c.%
\right]  +m_{n}c_{n}^{\dagger }\sigma _{z}c_{n}  \nonumber \\
&&+t_{-}\left[ c_{n}^{\dagger }\frac{\sigma _{y}}{2i}c_{n-\boldsymbol{a}%
_{2}}+c_{n}^{\dagger }\frac{\sigma _{z}}{2}c_{n-\boldsymbol{a}_{2}}+h.c.%
\right]     \nonumber \\
&&+m_{n-\boldsymbol{a}_{2}}c_{n-\boldsymbol{a}_{2}}^{\dagger }\sigma
_{z}c_{n-\boldsymbol{a}_{2}}, \label{mod2ban-r}
\end{eqnarray}
where $\boldsymbol{a}_{1} = (1,0)l$ and $\boldsymbol{a}_{2} = (0,1)l$. The on-site energy 
$m_{+}=m_{+,n} + m_{+,n-\boldsymbol{a}_{2}} $, $m_{-}=m_{-,n} + m_{-,n-\boldsymbol{a}_{2}} $, $m_{n} = m_{+,n} + m_{-,n}$ and $m_{n-\boldsymbol{a}_{2}} = m_{+,n-\boldsymbol{a}_{2}}
+ m_{-,n-\boldsymbol{a}_{2}}$. Here $m_{n-\boldsymbol{a}_{2}} = m_{n+\boldsymbol{a}_{2}} $ due to periodicity.

To study the edge states one needs to determine the Dirac points. Taking  $t_0 = t_+ = t_- =1$, the components of the Hamiltonian \eqref{mod2band-m} are $h_x = \sin k_x$, $h_y = H_{k_{y},\text{I}}  \sin k_{y} + H_{k_{y},\text{II}} \sin \left( -k_{y}\right)$ and $h_z = \cos k_{x} + H_{k_{y},\text{I}} \left( \cos k_{y}  + m_{+} \right)  + H_{k_{y},\text{II}} \left[ \cos \left(-k_{y}\right)  + m_{-} \right]$. Letting $h_x = h_y =0$, one solves out the Dirac points and the topological defects, as shown in Table \ref{Table-ModifiedModel}.
\begin{table}[H]
\caption{Monopole and meron defects at Dirac points.
\label{Table-ModifiedModel}}
\begin{center}
\begin{tabular}{ccc}
\hline
Dirac points $\left( k_x, k_y \right)$  & ~Monopoles ($h_z =0$) & ~Merons ($ h_z \neq 0$) \\
\hline
$\left( 0, 0 \right)$  & $m_+ = -2$ & $m_+ \neq -2$  \\
$\left( 0, \pi \right)$  & $m_+ = 0$ & $m_+ \neq 0$  \\
$\left( 0, 2\pi \right)$  & $m_- = -2$ & $m_- \neq -2$  \\
$\left( 0, 3\pi \right)$  & $m_- = 0$ & $m_- \neq 0 $  \\
$\left( \pi, 0 \right)$  & $m_+ = 0 $ & $m_+ \neq 0$  \\
$\left( \pi, \pi \right)$  & $m_+ = 2$ & $m_+ \neq 2$  \\
$\left( \pi, 2\pi \right)$  & $m_- = 0 $ & $m_- \neq 0 $  \\
$\left( \pi, 3\pi \right)$  & $m_- = 2$ & $m_- \neq 2$  \\
\hline
\end{tabular}
\end{center}
\end{table}

%%%%%%%%%%%%%%%%%%%%%%%%%%%%%%%%%%
%%%%%%%%%%%%%%%%%%%%%%%%%%%%%%%%%%

\textit{Discussion.}
%
%In this paper we propose a knot theory for topological insulators, and use it to re-derive the Pontrjagin index of a typical two-band model, eq.\eqref{eq:201908011}. By modifying eq.\eqref{eq:201908011} we propose possible models of high topological numbers, eqs.\eqref{modLinNum2-A-gammaAB} and \eqref{Linknum2-gammaA}--\eqref{mod2ban-r}.
%
It should be addressed that the separability condition \eqref{hkxky-Def} is a strong requirement, which confines the proposed method to be suitable only for square and rectangular lattices with NN interactions. Indeed, for a two-dimensional lattice in the real space, $  H=-t \sum_{ij,\alpha}\left(c^{\dagger}_{i\alpha}c_{j\alpha}+h.c.\right)$, with $c^{\dagger}_{i\alpha}$ and $c_{i\alpha}$ the creation and annihilation operators at Site $i$ with spin $\alpha$. %The $ ij $ denote the connections between neighboring sites ($\langle ij \rangle$ for NN, $\langle\langle ij \rangle\rangle$ for NNN, etc.).
[If magnetic couplings and spin-orbit (say, Rashba) couplings are involved, the Hamiltonian needs further modification.]
Under the Fourier transformation, $
\sum_{j}c_{j} =\sum_{\mathbf{k}}c_{\mathbf{k}}e^{i\mathbf{k}\cdot\boldsymbol{r}}$,
with $\boldsymbol{r}$ the vector connecting Site $i$ (origin) and Site $j$, we have $ \sum_{ij}c_{i}^{\dagger}c_{j}=
 \sum_{\mathbf{k}}c^{\dagger}_{\mathbf{k}}  \left(\sum_{\boldsymbol{r}} e^{i \mathbf{k}\cdot \boldsymbol{r}} \right) c_{\mathbf{k}}$ --- if the RHS can be separated into a pure $k_{x}$ part plus a pure $k_{y}$ part, the desired separation of eq.\eqref{hkxky-Def} is achievable. It strongly depends on the structure of the studied lattice; one can verify that only square and rectangular lattices meet this requirement, in which $k_x$ and $k_y$ lie in parallel to the vectors $\boldsymbol{a}_1$ and $\boldsymbol{a}_2$, respectively.  Without that condition the vector representation for $\left( k_x , k_y \right)$ fails to exist. 

Our next work is to study triangular and oblique lattices such as honeycomb, Kagome, etc., as well as NNN interactions. A promising way to overcome the separability difficulty is: first, to introduce a non-orthogonal decomposition $\left( \mathbf{k}_1 , \mathbf{k}_2 \right)$ for the momentum $\mathbf{k}$ to replace the orthogonal $\left( k_x , k_y \right)$, with $\mathbf{k}_1$ and $\mathbf{k}_2$ parallel to $\boldsymbol{a}_1$ and $\boldsymbol{a}_2$, respectively; second, to further turn the problem into the square/rectangular case by introducing the complex coordinates and conformal transformations.

\textit{Conclusion.} In this paper a knot theory for two-dimensional square lattice is developed. Our calculation reveals that the Pontrjagin topological index $P$ in a two-band model could be regarded as a Gauss linking number between knots, which leads to successful re-derivation of the typical evaluations of topological number $P = 0, \pm 1$ in literature. Furthermore, we propose a modified two-band model in order to achieve higher topological numbers, $P = 0, \pm 1, \pm 2$. The corresponding Hamiltonian, lattice structure and Dirac points are discussed as well.

Recently in the search of high performance devices many efforts were placed in designing high topological number material in multi-layer lattice \cite{10.1126/science.aax8156}. The significance of this paper lies in opening a new direction in this research beyond the previous ones, that is, to focus on mono-layer lattice and consider benefit of the intrinsic symmetry.

%%%%%%%%%%%%%%%%%%%%%%%%%%%%%%%%%%%%%%%%%%%%%%%%%%%%%%%%%%%%%%%%%%%
%%%%%%%%%%%%%%%%%%%% Acknowledgements %%%%%%%%%%%%%%%%%%%%%%%%%%%%%
%%%%%%%%%%%%%%%%%%%%%%%%%%%%%%%%%%%%%%%%%%%%%%%%%%%%%%%%%%%%%%%%%%%

\textit{Acknowledgements.} The authors wish to thank Prof. Wei LI
for useful discussions. XL and ZC acknowledge support from
the National Science Foundation of China No.11572005 and
the Natural Science Foundation of Beijing No.Z180007. WH acknowledges support from National Science Foundation of China No.11874003 and No.51672018.

\bibliography{reference.bib}

\end{document}